\documentclass{iaus}
\usepackage{graphicx}

\def\ltsim{\mathrel{\hbox{\rlap{\hbox{\lower4pt\hbox{$\sim$}}}\hbox{$<$}}}}

\title[The MiMeS Project: First Results] 
{The MiMeS Project: First Results}
\author[J.H. Grunhut et al.]{J.H. Grunhut$^1$, E. Alecian$^{1,2}$, D.A. Bohlender$^3$, J.-C. Bouret$^4$, J. Grunhut$^1$, H. Henrichs$^5$, C. Neiner$^6$, V. Petit$^7$, N. St. Louis$^8$, M. Auri\`ere$^9$, O. Kochukhov$^{10}$, J. Silvester$^1$, G.A. Wade$^1$, A. ud-Doula$^{11}$\\and the MiMeS Collaboration\thanks{www.physics.queensu.ca/$\sim$wade/mimes}}   
\affiliation{$^1$Royal Military College of Canada, $^2$LESIA, France, $^3$Canadian Astronomy Data Centre,$^4$LAM, France,$^5$Ast. Inst. Amsterdam, Netherlands, $^6$GEPI, France, $^7$Universit\'e Laval, Canada, $^8$Univ. de Montr\'eal, Canada, $^{9}$LAT, France, $^{10}$Uppsala University, Sweden, $^{11}$Morrisville State College, USA}    

\pubyear{2009}
\volume{259}  
\pagerange{100--100}
\date{"YOUR MAILING DATE"  and in revised form ??}
 \jname{Cosmic Magnetic Fields: From Planets,
to Stars and Galaxies} \editors{K.G. Strassmeier, A.G. Kosovichev \&
J.E. Beckman, eds.}

\begin{document}

\maketitle

\begin{abstract}
Massive stars are those stars with initial masses above about 8 times that of the sun, eventually leading to catastrophic explosions in the form of supernovae. These represent the most massive and luminous stellar component of the Universe, and are the crucibles in which the lion's share of the chemical elements are forged. These rapidly-evolving stars drive the chemistry, structure and evolution of galaxies, dominating the ecology of the Universe - not only as supernovae, but also during their entire lifetimes - with far-reaching consequences. 

Although the existence of magnetic fields in massive stars is no longer in question, our knowledge of the basic statistical properties of massive star magnetic fields is seriously incomplete. The Magnetism in Massive Stars (MiMeS) Project represents a comprehensive, multidisciplinary strategy by an international team of recognized researchers to address the Òbig questionsÓ related to the complex and puzzling magnetism of massive stars. This paper present the first results of the MiMeS Large Program at the Canada-France-Hawaii Telescope.
\end{abstract}

\firstsection 
\section{Introduction}
The Magnetism in Massive Stars (MiMeS) Project represents a comprehensive, multidisciplinary strategy by an international team of recognized researchers to address the Òbig questionsÓ related to the complex and puzzling magnetism of massive stars. Recently, MiMeS was awarded "Large Program" status by both Canada and France at the Canada-France-Hawaii Telescope (CFHT), where the Project has been allocated 640 hours of dedicated time with the ESPaDOnS spectropolarimeter from late 2008 through 2012. 

The structure of the MiMeS Large Program includes 255 hours for a 20-target "Targeted Component" (TC)â which will obtained time-resolved high-resolution spectropolarimetery of known magnetic massive stars in Stokes $I$ and $V$ and for some targets also Stokes $Q$ and $U$, in addition to 385 hours dedicated to the $\sim150$-target "Survey Component" (SC), with the goal to provide critical missing information about field incidence and statistical field properties of a much larger sample of massive stars. For more information about the goals and structure of the MiMeS Large Program, see Wade et al. (these proceedings).

\section{First Results}
MiMeS observations first began in August 2008, since which time over 200 polarised spectra of approximately 50 targets have been acquired. The signal-to-noise ratio of the spectra have been in good agreement with the MiMeS exposure model (see Wade et al., these proceedings), with 80\% of spectra achieving more than 0.8 times the predicted S/N. The precision of the magnetic diagnosis has also generally exceeded expectations.

Seven of the observed targets are TC stars, while the remainder are SC objects (primarily field O, B and Be stars). The first MiMeS observation, of the A0p star HD~170973, was also its first detection. Preliminary Least-Squares Deconvolved (LSD; Donati et al. 1997) Stokes $I$ and $V$ LSD profiles, as well as the $N$ diagnostic null profile, of HD~170973, along with those of other SC targets, are shown in Fig~1. The SC targets presented in Fig. 1 span much of the range of spectroscopic types, rotational velocities and emission properties of stars to be observed within the context of MiMeS, and provide a reasonable overview of the typical precision expected from the Survey Component. In addition to some $\sim 40$ observed SC targets in which no magnetic field is detected, one new magnetic massive star appears to have been identified. The physical characteristics of this new detection make it a fascinating and unique object for further detailed study. However, the annoucement of the details of this discovery must await a confirming observation to be acquired very soon (in December 2008 or January 2009).

The low rate of field detections was fully expected, based on the small number of magnetic massive stars reported in the literature. This suggests that the incidence of magnetic stars begins to decrease with increasing mass somewhere past $\sim 4-5~M_\odot$ (see Power et al. 2008).

Further results of the MiMeS SC are reported by Petit et al. (these proceedings).

\begin{figure}
\centering
\includegraphics[width=2.2in]{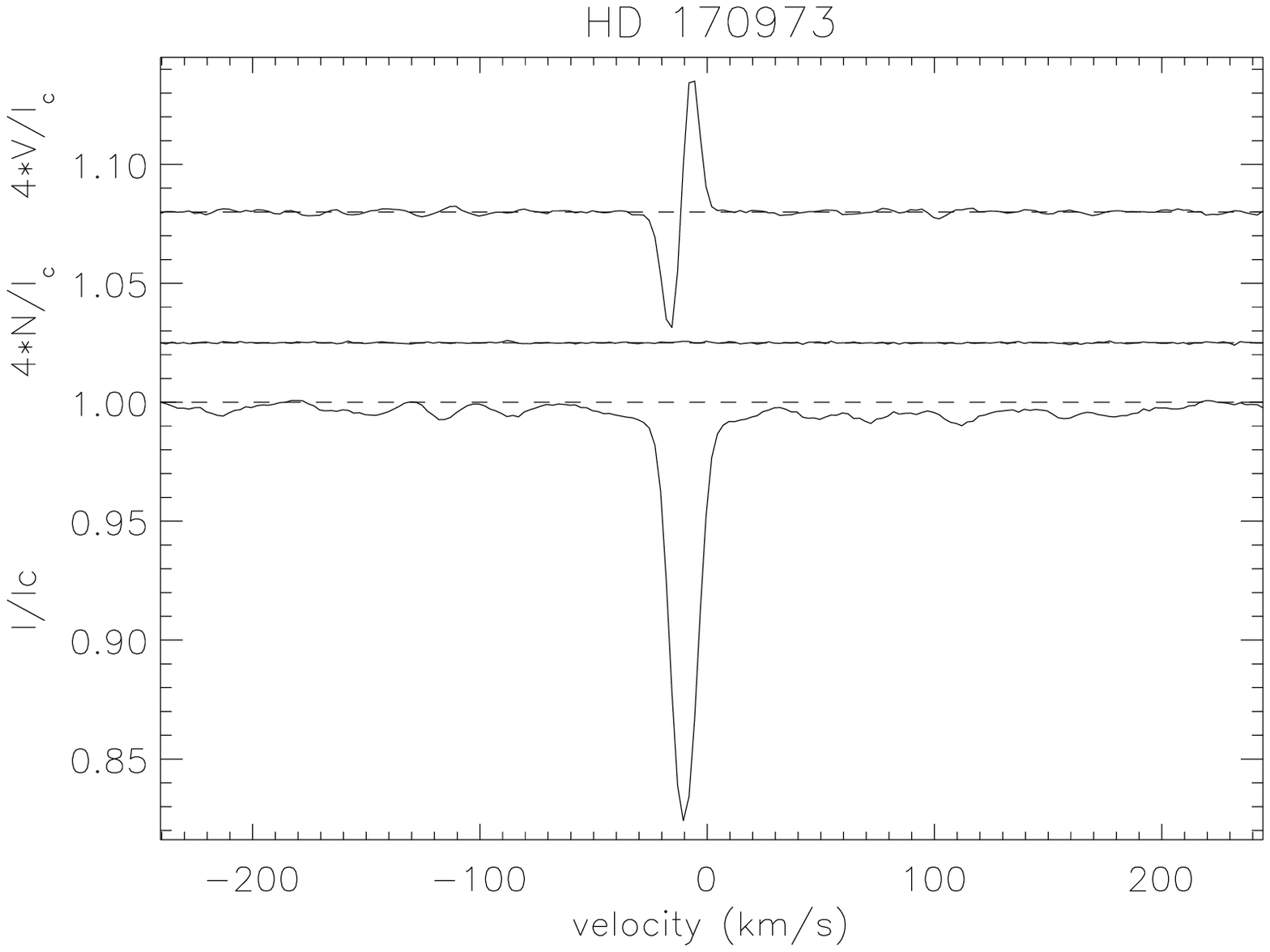}
\includegraphics[width=2.2in]{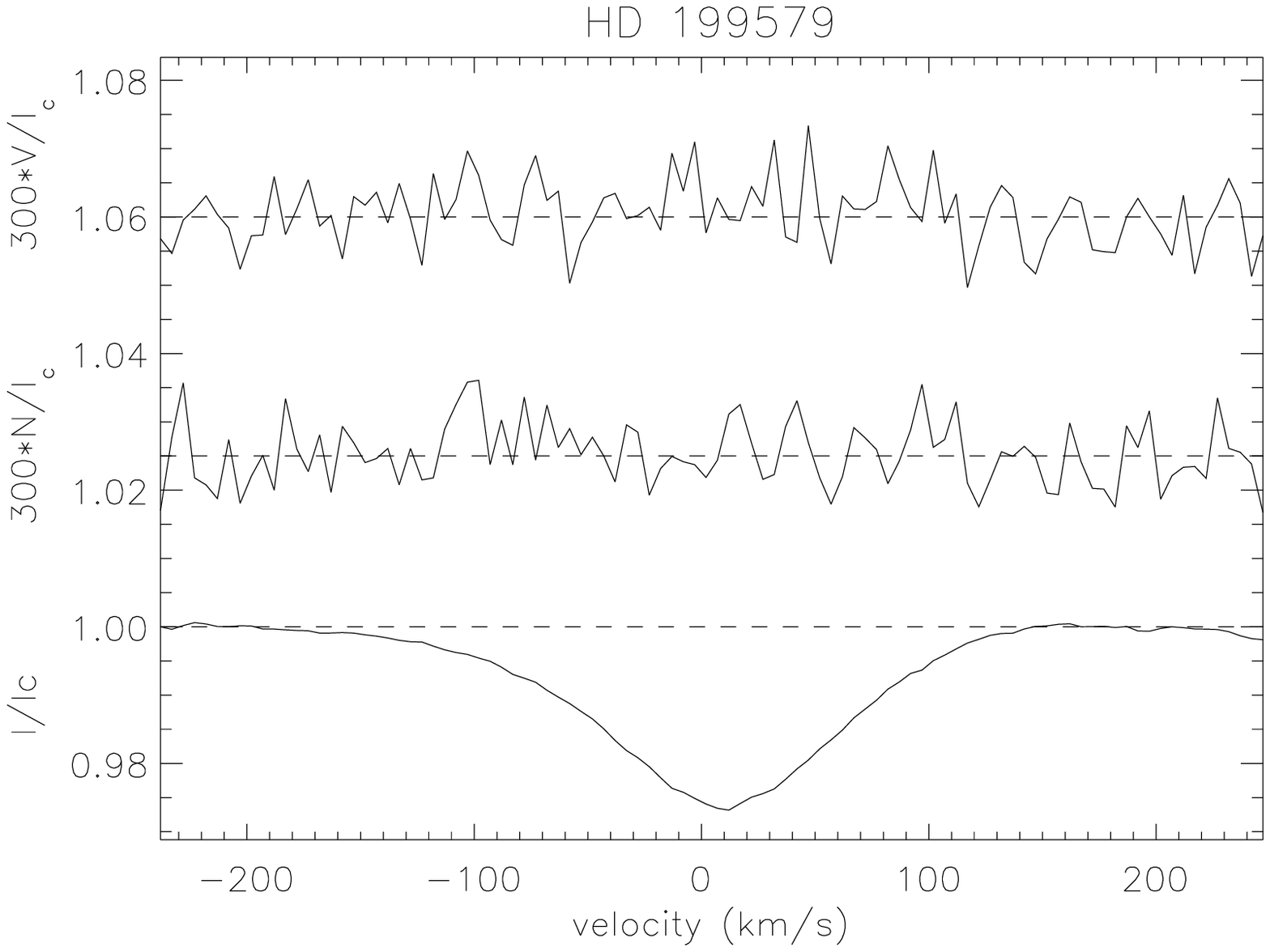}\\
\includegraphics[width=2.2in]{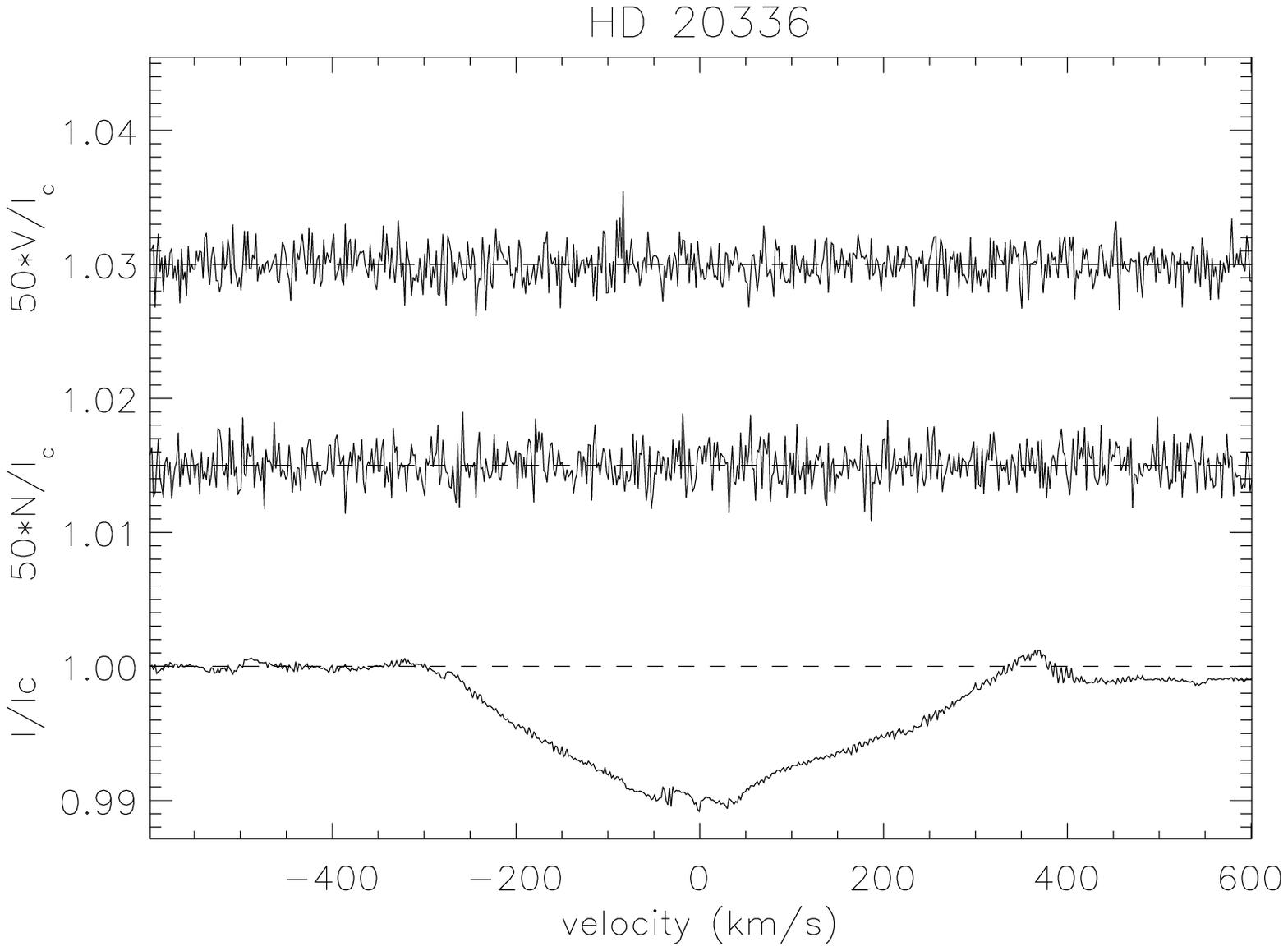}
\includegraphics[width=2.2in]{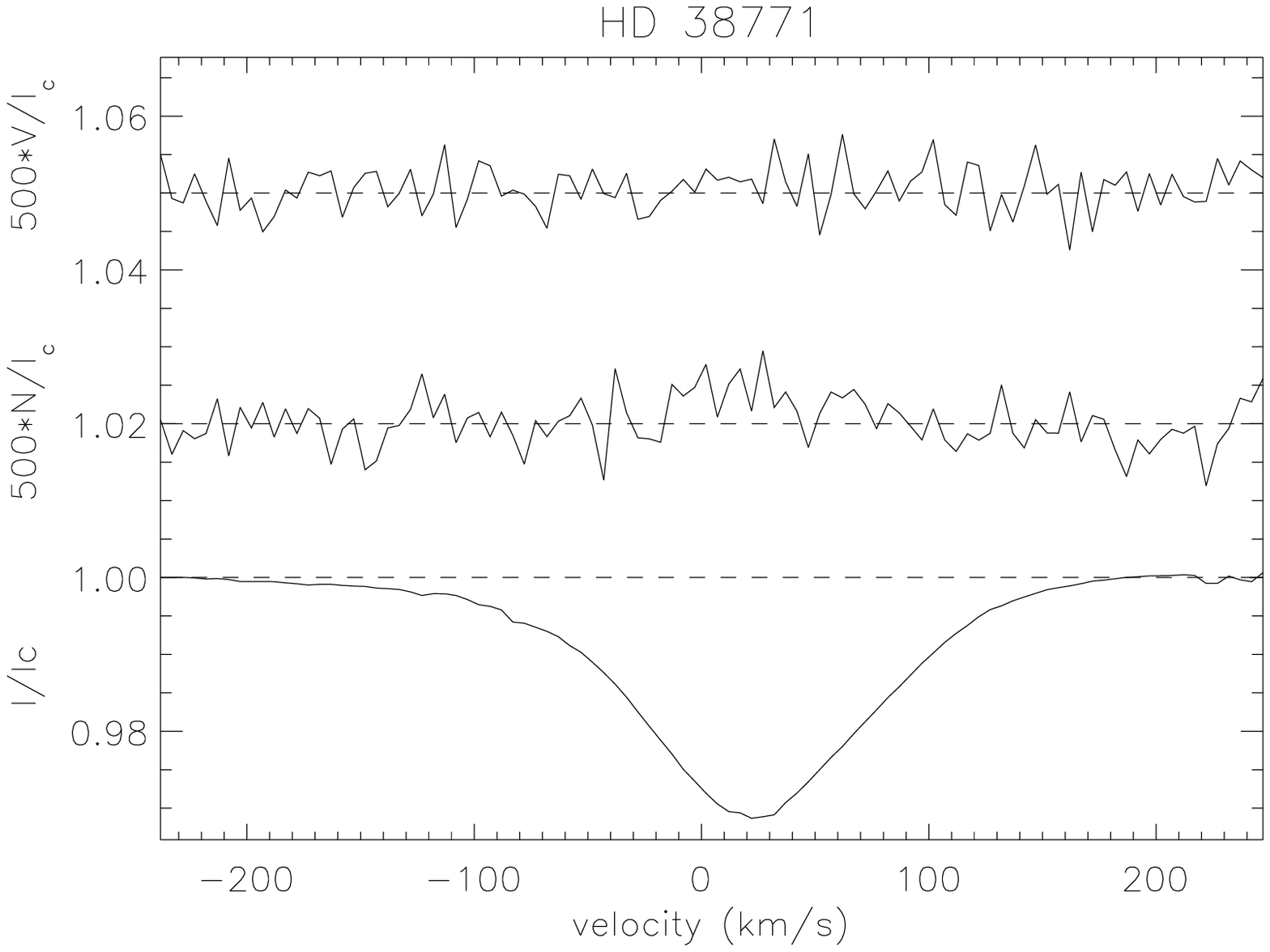}
\caption{Least-Squares Deconvolved (LSD) profiles of four MiMeS Survey Component targets observed during the first semester of observations. {\bf Top left:} HD~170973 (A0p), $v\sin i\simeq 20$~km\,s$^{-1}$, $B_\ell=-538\pm 11$~G (Stokes $V$ magnetic signature detected, dipole strength $B_{\rm d}>1.6$~kG); {\bf Top right:} HD~199579 (06Ve), $v\sin i\simeq 75$~km\,s$^{-1}$, $B_\ell=1\pm 29$~G (no detection, $B_{\rm d}\ltsim 100$~G); {\bf Bottom left:} HD~20336 (B2.5Vne), $v\sin i\simeq 175$~km\,s$^{-1}$, $B_\ell=32\pm 63$~G (no detection, $B_{\rm d}\ltsim 200$~G; {\bf Bottom right:} HD~38771 (B01ab), $v\sin i\simeq 65$~km\,s$^{-1}$, $B_\ell=-1\pm 8$~G (no detection, $B_{\rm d}\ltsim 30$~G). Each frame shows Stokes $I$ (bottom), scaled Stokes $V$ (top) and $N$ diagnostic null (middle). Note the large range of different scaling factors.}
\label{lsd_figs}
\end{figure}

\end{document}